\documentclass[a4paper,fleqn,12pt]{cas-sc}
\usepackage{soul}
\usepackage[sort&compress,numbers]{natbib}
\usepackage{lineno}
\usepackage{setspace}

\def\tsc#1{\csdef{#1}{\textsc{\lowercase{#1}}\xspace}}
\tsc{WGM}
\tsc{QE}
\tsc{EP}
\tsc{PMS}
\tsc{BEC}
\tsc{DE}

\begin{document}
\let\WriteBookmarks\relax
\def\floatpagepagefraction{1}
\def\textpagefraction{.001}
\shorttitle{Charge Localization and Hopping in a Topologically Engineered GNR}
\shortauthors{Pereira J\'unior \textit{et~al}.}

\title [mode = title]{Charge Localization and Hopping in a Topologically Engineered Graphene Nanoribbon}

\author[1]{ Marcelo Lopes Pereira J\'unior}
\author[1]{ Pedro Henrique de Oliveira Neto}
\author[1]{ Dem\'etrio Ant\^onio da Silva Filho}
\author[2]{ Leonardo Evaristo de Sousa}
\author[1]{ Geraldo Magela e Silva}
\author[1]{ Luiz Ant\^onio Ribeiro J\'unior}

\cormark[1]
\ead{ribeirojr@unb.br}
\cortext[cor1]{Corresponding author}

\address[1]{Institute of Physics, University of Bras\'ilia, 70910-900, Bras\'ilia, Brazil.}
\address[2]{Theoretical and Structural Chemistry Group, State University of Goi\'as, Anapolis, 75132-903, Goi\'as, Brazil.}

\begin{abstract}
Graphene nanoribbons (GNRs) are promising two-dimensional materials with various technological applications, in particular for the armchair GNR families that have a semiconductor character. Recently, methods that allowed for the control of GNR's topology have been developed, resulting in the production of nanoribbons composed of alternating segments of two distinct armchair GNR families (7 and 9-AGNRs) connected in heterojunctions. This GNR displays two topological bands that lie between the valence and conduction bands that effectively modulates the nanoribbon bandgap. Here, we employ a two-dimensional extension of the Su-Schrieffer-Heeger model to study morphological and electronic properties of this new material in both neutral and charged states. Results demonstrate that charge injection in this system results in the formation of polarons that localize strictly in the 9-AGNRs segments of the system and whose mobility is highly impaired by the system's topology. We further show polaron displacement by means of hopping between 9-AGNR portions of the system, suggesting this mechanism for charge transport in this material.
\end{abstract}





\maketitle
\doublespacing

\section{Introduction}

Optoelectronic devices are responsible for the capture, control, and emission of light and are widely used nowadays~\cite{sirringhaus1998,li2006nanowire,bhattacharya1997semiconductor}. The most common materials used in the production of these devices are inorganic~\cite{shaw2001organic, mitzi2001organic}. However, ever since the discovery of conductive organic crystals in the 1960s ~\cite{pron2002processible}, the search for the production and improvement of devices manufactured from organic materials has been drawing much attention from the scientific and industrial communities, as these devices offer lower manufacturing costs due to the abundance of raw material, their mechanical flexibility, the possibility of making thin, light and even transparent devices, as well as little environmental impact ~\cite{burroughes1990light,forrest2004path, pichler2005encapsulation,logothetidis2008flexible, forrest2004path,mohanty2012nanotomy}. Among the several applications that can be manufactured with organic electronics, the most common are organic light-emitting diodes (OLED) ~\cite{burroughes1990light, noh2020synthesis} and organic photovoltaic (OPV) devices ~\cite{yu1995polymer,burschka2013sequential,erden2020synthesis} and also organic field-effect transistors (OFET) ~\cite{friend1999electroluminescence,dimitrakopoulos2002organic}.

Among the various classes of organic materials currently under study, graphene deserves a place of prominence ~\cite{stankovich2006graphene,novoselov2007rise}. Synthesized for the first time in 2004 ~\cite{novoselov2004electric}, it consists in a two-dimensional carbon honeycomb lattice and displays several physical properties of great potential for use in new technologies ~\cite{novoselov2012roadmap, neto2009electronic, ferrari2006raman}. Graphene sheets, however, have a zero bandgap ~\cite{garcia2011group, singh2011graphene}, which prevents them from being used in a semiconductor capacity. This problem has been overcome with the production of graphene nanoribbons (GNR), which are long strips with widths up to 50 nm that may actually present non-zero bandgaps~\cite{barone2006electronic, chen2007graphene}. These nanoribbons differ by their edge types~\cite{zhang2012experimentally}, with the most common being: armchair GNR (AGNR) and zigzag GNR (ZGNR)~\cite{li2008chemically,barone2006electronic}. AGNRs, in turn, are divided into three families, $ n=3p $, $ n=3p+1 $, and $ n=3p+2 $, where $p$ is an integer and $n$ is the number of carbon atoms along the width of the nanoribbon. The AGNRs of the $ 3p $ and $ 3p+1 $ families have know to display semiconductor behavior~\cite{da2016polaron,ribeiro2015transport}.

Recently, a procedure for topological band engineering of GNRs was reported by means of the creation of a heterojunction between GNRs of two different families~\cite{rizzo2018topological}. By creating this structure that alternates between a $n=7$ AGNR and a $n=9$ AGNR with a non-trivial topology (7,9-AGNR), they demonstrated the appearance of two topological bands between the valence and conduction bands, demonstrating how the capacity of precisely controlling electronic topology allows the tuning of the system's bandgap. Since only electronic properties of this new nanorribon have been described, this important result raises the question of how such topological changes affect charge transport in this material. Importantly, the influence of electronic correlations on the topological states of 7,9-AGNR heterostructures on Au(111) was theoretically investigated by using a GW approach combined with an effective Hubbard Hamiltonian \cite{jauho_NL}. Through this combined approach, the results have shown that strong local electronic correlations are present in both the edges of the nanoribbon. Polarons and bipolarons take place in organic systems due to lattice relaxation effects \cite{heeger_ACI,pereirajr_2017a,pereirajr_jmcc_2020,abreu_jmm,pereirajr_sm,pereirajr_jmm_2019a,pereirajr_2019_jpcc,pereirajr_2020_pccp}. To account for such effects, the electron-phonon coupling term should be considered. This feature, absent in the study conducted in reference \cite{jauho_NL}. In this sense, other approaches are required to describe the presence of polarons and bipolarons in organic-based lattices.   

In semiconductor AGNRs, the electronic properties of the system are substantially altered by the deformations of the lattice sites, as the displacements of atoms and electrons is coupled and responds to external factors in a connected manner. As a result, charge injection to these systems produces quasi-particles such as the polaron, which corresponds to an electron or hole coupled to the lattice deformations that appear due to polarization of the system in the region of excess charge and play the role of charge carriers in these materials. In this work, the electronic and morphological properties of this recently synthesized 7,9-AGNR are studied in both neutral and charged states by means of a two-dimensional extension of the Su-Schrieffer-Heeger (SSH) model. Lattice dynamics is investigated within an Ehrenfest molecular dynamics approach. Results demonstrate that charge injection in this system results in polaron formation. This quasi-particle is seen to localize strictly in the 9-AGNR segments of the system and, for a range of electron-phonon coupling, it moves under the influence of an external electric field by means of a hopping mechanism between the 9-AGNRs portions of the system. 

\section{Methodology}

To study the transport of quasi-particles in hybrid structures formed by the heterojunction of AGNRs with widths of 7 ($ 3p+1 $ family) and 9 ($ 3p $ family), we used a SSH Hamiltonian model, in which the electronic part of the system is described quantum mechanically while the lattice part is treated classically. The two parts of the Hamiltonian are connected by an electron-phonon coupling term that is used to include lattice relaxation to a two-dimensional tight-binding model. Since the position of atoms in graphene nanoribbons is not substantially altered, the electronic transfer integrals for $\pi$ electrons can be expanded in first order ~\cite{da2016polaron}. As such, the hopping term is given by
\begin{equation}
\label{hop}
t_{i,j} = t_0 - \alpha \eta_{i,j}
\end{equation}
where $ t_0 $ is the hopping integral of the system with all atoms equally spaced, $ \alpha $ represents the electron-phonon coupling that is responsible for the interaction between the electronic and lattice degrees of freedom, and $ \eta_{i,j} $ are the variations in the bond-lengths of two neighboring sites $i$ and $j$.

\begin{figure}
	\centering
	\includegraphics[scale=1.0]{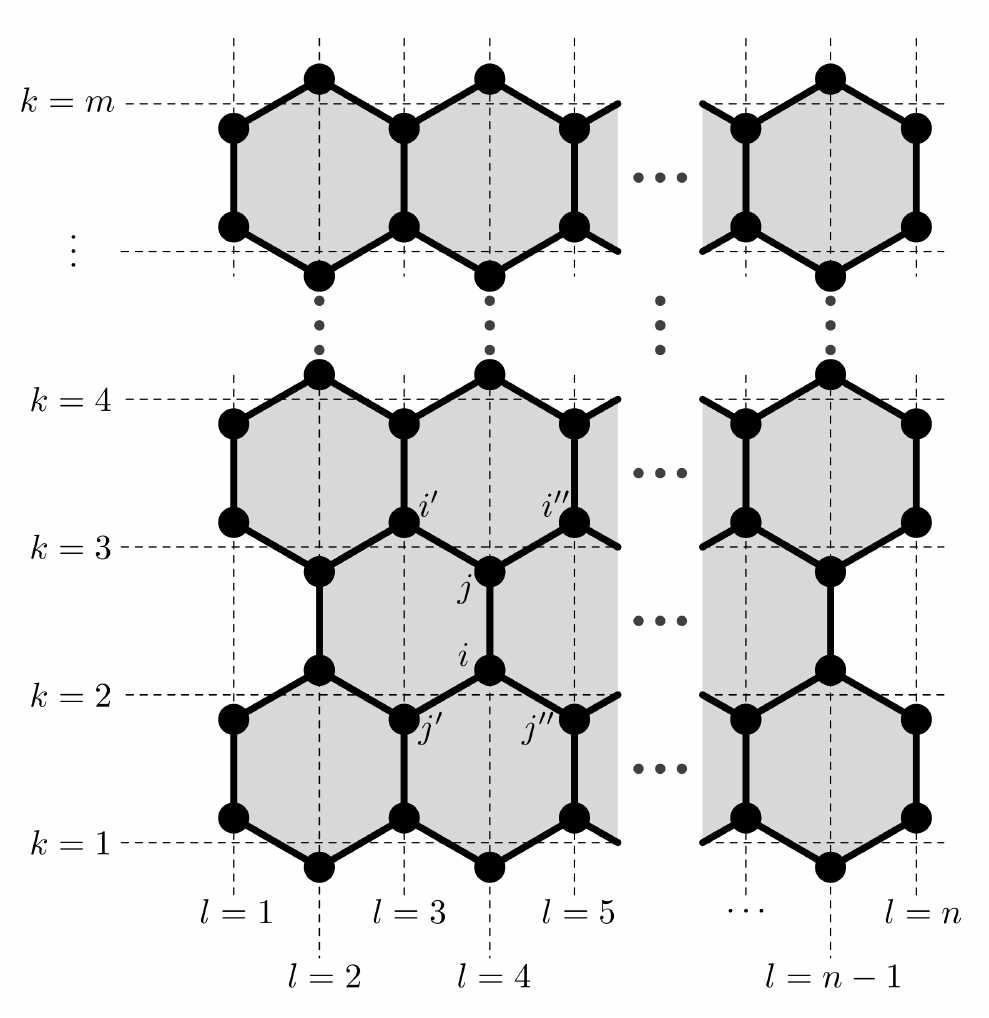}
	\caption{Schematic representation of indices in armchair graphene nanoribbons.
	}
	\label{fig1}
\end{figure}

The Hamiltonian model used here is given by the expression
\begin{equation}
\begin{split}
H &= -\sum_{\langle i,j \rangle, s} \left(t_{i,j}^{\phantom{*}} C_{i,s}^\dag C_{j,s}^{\phantom{\dag}}+t_{i,j}^* C_{j,s}^\dag C_{i,s}^{\phantom{\dag}} \right ) \\
&\hspace{0.1cm}+ \frac12K\sum_{\langle i,j \rangle} \eta_{i,j}^2+\frac1{2M}\sum_ip_i^2
\end{split}
\end{equation}
where $ \langle i,j \rangle $ represents the indexes of neighboring sites (see Figure \ref{fig1}), $C_{i,s}^{\phantom{\dag}} $ is the $\pi$-electron annihilation operator on site $i$ with spin $s$ and $ C_{i,s}^{\dag} $ represents the corresponding creation operator. The second term is the effective potential associated with sigma bonds between carbon atoms, modeled according to the harmonic approximation with $K$ being the elastic constant. The last term describes the kinetic energy of the sites in terms of their momenta $ p_i $ and mass $M$.

The values for the different model parameters used are 2.7 eV for $t_0$ and 21 eV/\AA$^2$ for $K$. Values for $ \alpha $ ranged from  0.1 eV/\AA~ to 6.0 eV/\AA. These choices of model parameters follow other theoretical and experimental works~\cite{barone2006electronic, de2012electron, ribeiro2015transport, kotov2012electron,yan2007electric, neto2009c, yan2007raman}. 

Starting the iteration from an initial set of coordinates $ \{\eta_{i,j}\} $, a self-consistent stationary solution (with $ p_i = 0 $) of the system is determined~\cite{lima2006dynamical}. The ground state is obtained with the diagonalization of the electronic Hamiltonian, according to the expression
\begin{equation}
H=-\sum_{k,s} E_k a^\dag_{k,s} a_{k,s}^{\phantom{\dag}},
\label{eq3}
\end{equation}
where $ E_k $ are the eigenenergies of the electronic system. To do this procedure, it is necessary to obtain the operators ${a_{k,s}}$, which enable a diagonal Hamiltonian. These operators are obtained in LCAO form,
\begin{equation}
a_{k,s}=\sum_i \psi_{k,i,s} C_{i,s}.
\end{equation}
From these considerations, the electronic Hamiltonian becomes
\begin{equation}
H = -\sum_{\langle i,j \rangle,s,k,k'} \left( t_{i,j}\psi^{\phantom{*}}_{k,i,s}\psi^*_{k',j,s} + t^*_{i,j}\psi^{\phantom{*}}_{k,j,s}\psi^*_{k',i,s}\right )a^\dag_{k,s} a^{\phantom\dag}_{k',s}
\end{equation}
which is diagonalized and becomes Equation \ref{eq3} as long as the condition (Equation \ref{eq6}) is satisfied
\begin{equation}
-t_{i,j}\psi_{k,j,s}-t_{i,j'}\psi_{k,j',s}-t_{i,j''}\psi_{k,j'',s} = E_k\psi_{k,i,s},
\label{eq6}
\end{equation}
is satisfied for neighboring sites $i$, $j$; $i$,$j'$; and $i$,$j''$ (see Figure \ref{fig1}). The result of the procedure of diagonalization is the energies of the electronic states and the wave functions for the ground state.

The concomitant self-consistent lattice solution is obtained from the Euler-Lagrange equations:
\begin{equation}
\frac d{dt}\left( \frac{\partial\langle L \rangle}{\partial \dot \xi_l} \right ) - \frac{\partial \langle L \rangle}{\partial \xi_l} = 0,
\end{equation}
wherein the static case is
\begin{equation}
\frac{\partial \langle L \rangle}{\partial \xi_{i,j}} = 0.
\end{equation}
Thus, to take into account lattice effects, it is necessary to obtain the expectation value of the Lagrangean system, $ \langle \psi | L | \psi \rangle $, where $ | \psi \rangle $ is the Slater state represented in the second quantization formalism by $ | \psi \rangle = a_1^\dag a_2^\dag \cdots a_n^\dag | \; \rangle $. As such,
\begin{equation}
\begin{split}
L & = \frac{M}{2}\sum_l \dot\xi_l^2-\frac12K\sum_{\langle i,j \rangle} \eta_{i,j}^2                                                                                          \\
& \vspace{.1cm}+ \sum_{\langle i,j \rangle, s} \left(t_0 - \alpha \eta_{i,j} \right )\left(C_{i,s}^\dag C_{j,s}^{\phantom\dag} + C_{j,s}^\dag C_{i,s}^{\phantom\dag} \right ),
\end{split}
\end{equation}
thus,
\begin{equation}
\begin{split}
\langle L \rangle & = \frac{M}{2}\sum_i \dot\xi_i^2-\frac12K\sum_{\langle i,j \rangle}\eta_{i,j}^2  \\
& + \sum_{\langle i,j \rangle, s} \left( t_0-\alpha \eta_{i,j} \right)\left(B_{i,j}+B^*_{i,j}\right);
\end{split}
\end{equation}
with,
\begin{equation}
B_{i,j} \equiv \sum_{k,s}{'}\psi^*_{k,s}(i,t) \psi^{\phantom{*}}_{k,s}(j,t),
\end{equation}
where the sum is realized only for the occupied states. Note also that the last equation is responsible for the connection between the electronic and lattice parts of the system.

Thus, an initial set of coordinates $\{ \eta_{i,j} \}$ is used to start an auto-consistent calculation, where a corresponding electronic set $\{ \psi_{k,i,s} \}$ is obtained, which when solved for the lattice returns a new set of coordinates $\{ \eta_{i,j} \}$. The process is repeated until a given convergence criterion is satisfied. From the stationary solution $\{ \eta_{i,j} \}$ and $\{\psi_{k,i,s}\}$, the evolution of the system over time is performed by means of the time-dependent Schrödinger equation for electrons along with the solution of the Euler-Lagrange equation for the movement of atoms \cite{pereirajr_jpcl,pereirajr_jpcc_2019b}. Thus, the electronic time evolution is given by
\begin{equation}
|\psi_k(t+dt)\rangle = e^{-\frac i\hbar H(t) dt}|\psi_k(t)\rangle.
\end{equation}

Expanding the ket $ |\psi_k(t)\rangle $ in a basis of eigenstates of the electronic Hamiltonian at a given time $t$, we obtain
\begin{equation}
|\psi_k\rangle = \sum_l D_{k,l}|\phi_l(t)\rangle,
\end{equation}
and finally, we obtain the temporal evolution of the electronic part of the system, according to the final expression
\begin{equation}
|\psi_k(t+dt)\rangle = \sum_l \langle \phi_l(t)|\psi_k(t)\rangle e^{-i\varepsilon_l dt/\hbar}|\phi_l(t)\rangle.
\end{equation}

Or, in terms of wave functions
\begin{equation}
\psi_{k,s}(i,t+dt) = \sum_{l,m} \phi^*_{l,s}(m,t)\psi_{k,s}(m,t)e^{-i\varepsilon_l dt/\hbar}\phi_{l,s}(i,t),
\end{equation}
where $\{\phi_l\}$ and $\{\varepsilon_l\}$ are the eigenfunction and eigenvalues of the electronic Hamiltonian, respectively. The numerical integration of the last equation is performed as usual and has already been reported in our previous work ~\cite{lima2006dynamical}.

For the classical treatment governing the lattice part of the system, the complete Euler-Lagrange equations are required. Its solution can be written as a Newtonian equation able to describe the movements of the sites in the system and is given by
\begin{equation}
\begin{split}
M\ddot \eta_{i,j} &= \frac12K\left(\eta_{i,i'}+\eta_{i,i''}+\eta_{j,j'}+\eta_{j,j''} \right )-2K\eta_{i,j}\\ 
&+ \frac12\alpha\left(B_{i,i'}+B_{i,i''}+B_{j,j'}+B_{j,j''}-4B_{i,j} + \mathrm{c.c.}\right ).
\end{split}
\end{equation}

To perform the quasiparticle dynamics in the system, an external electric field, $\mathrm{\textbf{E}}(t)$, was included in our model. Here, this is done by inserting a time-dependent vector potential, $\mathrm{\textbf{A}}(t)$, through a Peierls Substitution for the electronic transfer integrals of the system, making the hopping term
\begin{equation}
t_{i,j} = e^{-i\gamma\mathrm{\textbf{A}}}\left(t_0 - \alpha \eta_{i,j} \right ),
\end{equation}
where $ \gamma \equiv ea/(\hbar c) $, with $a$ being the lattice parameter ($ a = 1.42 $ \AA{} in graphene nanoribbons), $e$ being the absolute value of the electronic charge, and $c$ the speed of light. The relationship between the time-dependent electric field and the potential vector is given by $ \mathrm{\textbf{E}}(t) = -(1/c)\dot{\mathrm{\textbf{A}}}(t) $. In our model, the electric field is activated adiabatically to avoid numerical oscillations that appear when the electric field is turned on abruptly~\cite{da2016polaron}.

\section{Results and Discuss}

\begin{figure*}
	\centering
	\includegraphics[width = 0.8\linewidth]{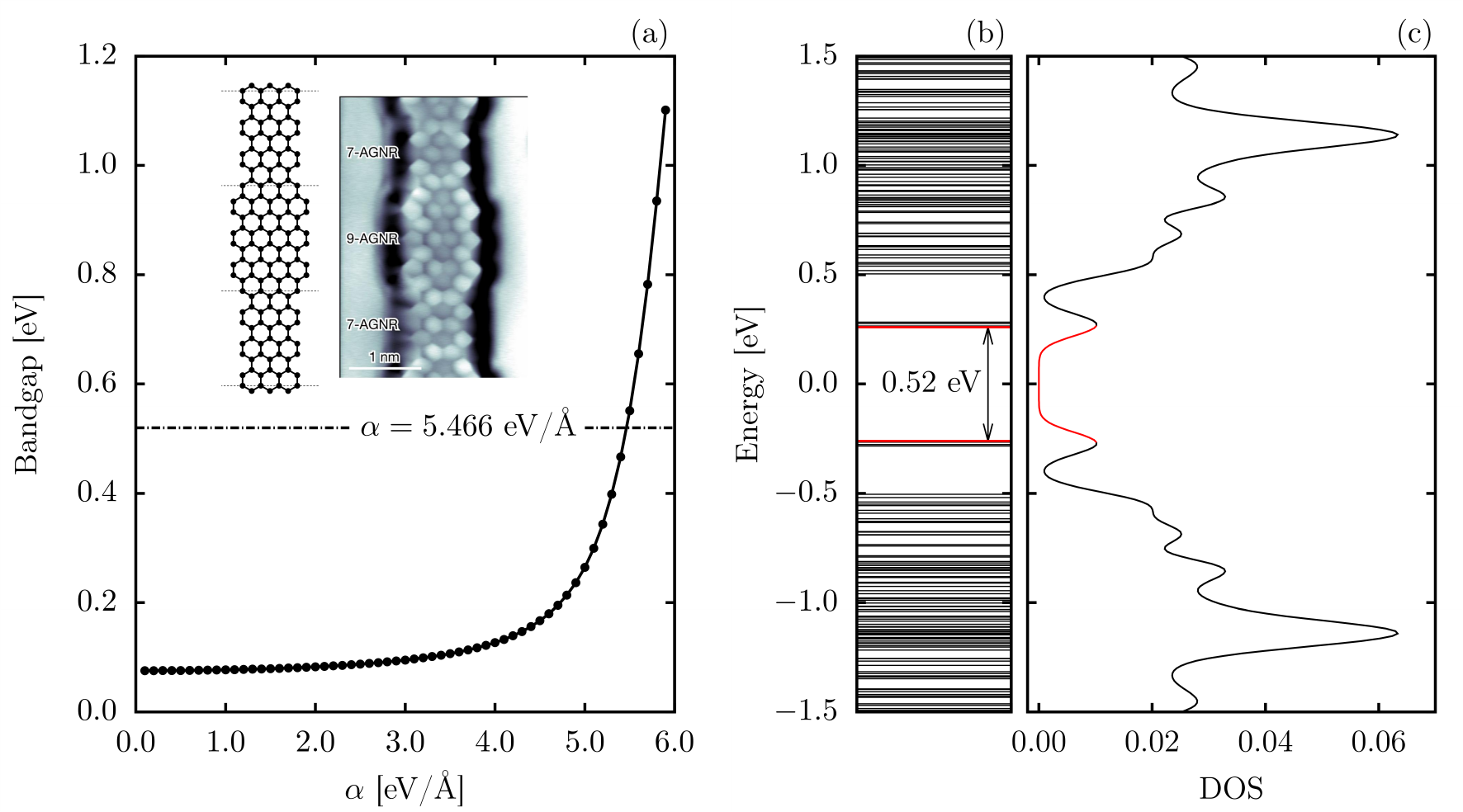}
	\caption{(a) Bandgap as a function of the electron-phonon coupling for the 7,9-GNR. The left internal panel presents a schematic representation of the studied structure and the right internal panel presents a synthesized fragment of 7,9-AGNR (ref. ~\cite{rizzo2018topological}), (b) energy levels and (c) density of states for the 7,9-AGNR with $\alpha=5.466 $ eV/\AA.}
	\label{fig2}
\end{figure*}

The structure of the graphene nanoribbon employed in the simulations is shown in the inset of Figure \ref{fig2}. This nanoribbon is composed of alternating segments of 7-AGNRs and 9-AGNRs linked in heterojuntions. For the simulations, nanoribbons with total length of 192 atoms were used. To be able to study the electronic properties of this nanoribbon, it is necessary to gauge the electron-phonon coupling ($\alpha$) that best characterizes it. The intensity of this coupling affects the resulting bandgap, as shown in Figure \ref{fig2}(a), in which bandgaps for the neutral 7,9-AGNR are determined as a function of $\alpha$. For couplings below 4.0 eV/\AA~, only slight changes in bandgap are observed with energy variations barely surpassing the 0.1 eV mark. For larger couplings, however, the bandgap is seen to become much more sensitive, rapidly increasing as $\alpha$ reaches 6.0 eV/\AA. Agreement with predicted bandgap or previous density functional theory and tight-binding calculations~\cite{rizzo2018topological} is achieved for $\alpha=5.466 $ eV/\AA, which produces the 0.52 eV bandgap seen in Figure \ref{fig2}(b). This energy gap corresponds to the energetic difference between the two topological bands that are marked in red in Figure \ref{fig2}(b) which are also in agreement with experimental results.

\begin{figure*}
	\centering
	\includegraphics[width=0.8\linewidth]{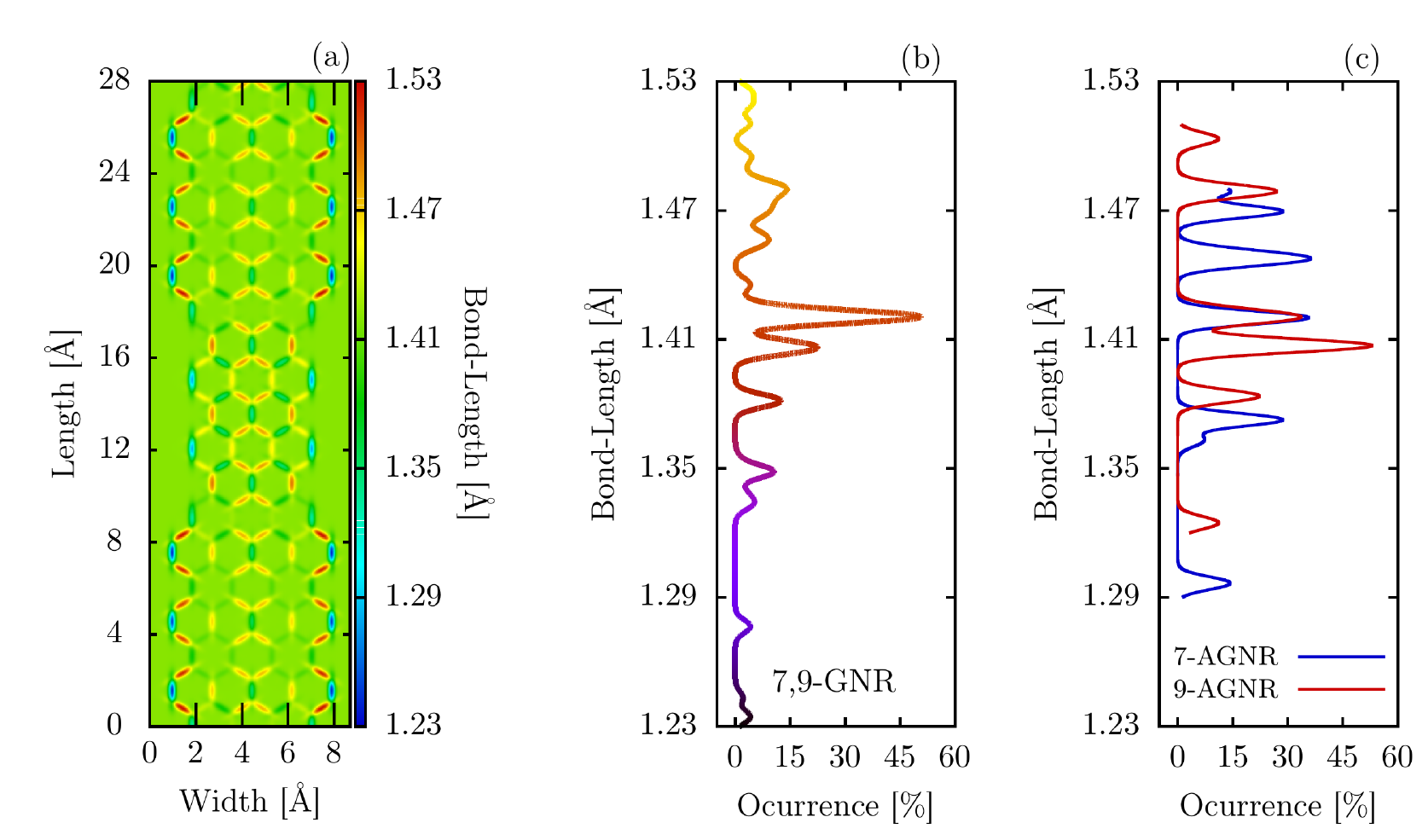}
	\caption{(a) Bond-lengths in a 7,9-GNR fragment; (b) bond-length distribution along the entire length of 7,9-GNR and, (c) bond-length distribution for the separate 7-AGNR and 9-AGNR.}
	\label{fig3}
\end{figure*}

From a morphological point of view, the 7,9-AGNR can be analyzed by looking into how its bond lengths are modified with respect to the bond length of a graphene sheet (1.42 \AA). These variations in bond lengths can be seen in Figure \ref{fig3}(a), in which hot and cold colors correspond, respectively, to the stretching and compression of bond lengths. It can be seen that edge bonds alternate between expansion and compression where the amplitude of such variation is larger in the 9-AGNR portion of the system. The middle carbon rings are seen to present slightly stretched bonds, but are kept closer together by contracted bonds in the vertical direction in Figure \ref{fig3}(a). Another feature that differentiates the 7 and 9-AGNR fragments is the presence of four aromatic rings in the 9-AGNR portion, as evidenced by their homogeneity~\cite{martin2012electronic}. The overall distribution of bond lengths in the 7,9-AGNR can be seen in Figure \ref{fig3}(b). This histogram shows that stretched bond lengths around 1.42 \AA~ are the most common in 7,9-AGNRs, a common feature of low bandgap GRNs. Around 10\% of bonds are stretched to 1.48 \AA, corresponding mostly to edge bonds. Compensating these enlarged bonds, contracted bonds are also found divided into well separated peaks around 1.4 \AA, 1.38 \AA~ and 1.35 \AA.   


\begin{figure*}
	\centering
	\includegraphics[width = \linewidth]{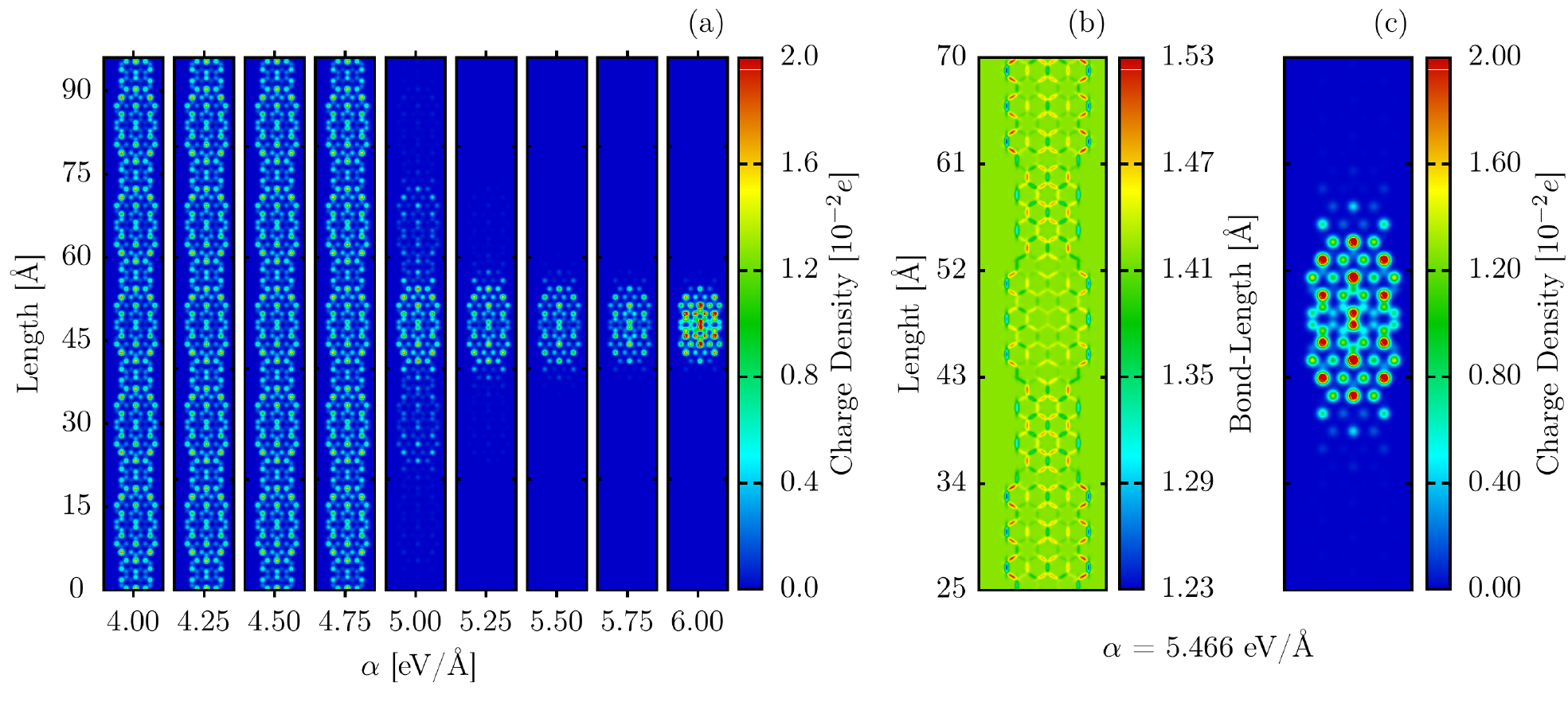}
	\caption{Atomic charge distribution for different electron-phonon coupling intensities (a), extended view of the bond-length (b) and charge density (c) in the presence of a polaron and with $\alpha = 5.466$ eV/\AA. The fragment presented corresponds to the zoom in the region of charge concentration of the nanoribbon with 288 \AA{} of length.}
	\label{fig4}
\end{figure*}

The more interesting aspects of the 7,9-AGNR heterojunction can be seen when a hole is injected in the system. The behavior of the excess charge is dependent on the intensity of the electron-phonon coupling. This can be seen in Figure \ref{fig4}(a), that presents a charge density plot for different $\alpha$. For $\alpha < 4.75$ eV/\AA, the excess charge is delocalized over the entire nanoribbon. As $\alpha$ grows larger, the excess charge becomes progressively more localized. The excess charge polarizes the structure and distorts the lattice, as it can be seen in Figure \ref{fig4}(b) for the particular case of $\alpha = 5.0$ eV/\AA. The largest differences in morphology are observed in a 9-AGNR fragment, between 25 and 35 \AA. Comparison with the neutral case (Figure \ref{fig3}(a)) or with other 9-AGNR portions reveals that bond length changing in this region now extends to the once aromatic rings. Charge distribution is observed to be symmetric with respect to both nanoribbon axes and localized within the 9-AGNR segment. This combination of charge accumulation and localized bond length distortion is a feature of the electron-phonon coupling and characterizes, in this case, a polaron. From an electronic standpoint, the presence of a polaron is confirmed by the appearance of two intragap energy levels. In the case of the 7,9-AGNR, these intragap levels lies between the two topological bands produced by the heterojuntion.   

A remarkable feature of polaron formation in 7,9-AGRNs is the fact that regardless of the set of initial coordinates employed in the simulation, charge accumulation always takes place in a 9-AGNR portion of the system. This is so even though both 7 and 9-AGNR, as members of the $3p + 1$ and $3p$ AGNR families, are known to be prone to polaron formation~\cite{kimouche2015ultra}. We conjecture that this behavior is due to the fact that the distortion of aromatic bonds found only in the 9-AGNR segments constitute significantly contribution to entropy increase in the nanoribbon, making the localization of polarons in these regions a process that minimizes the free energy of the system. 

\begin{figure}
	\centering
	\includegraphics[scale=0.85]{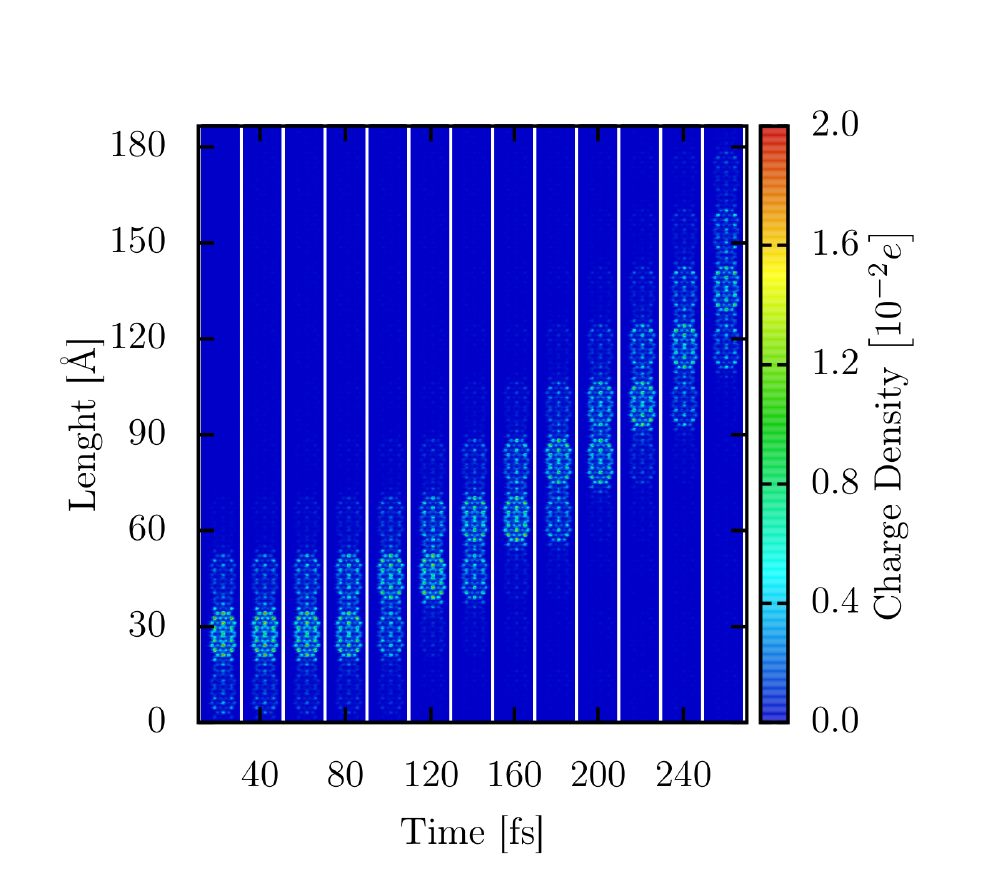}
	\includegraphics[scale=0.7]{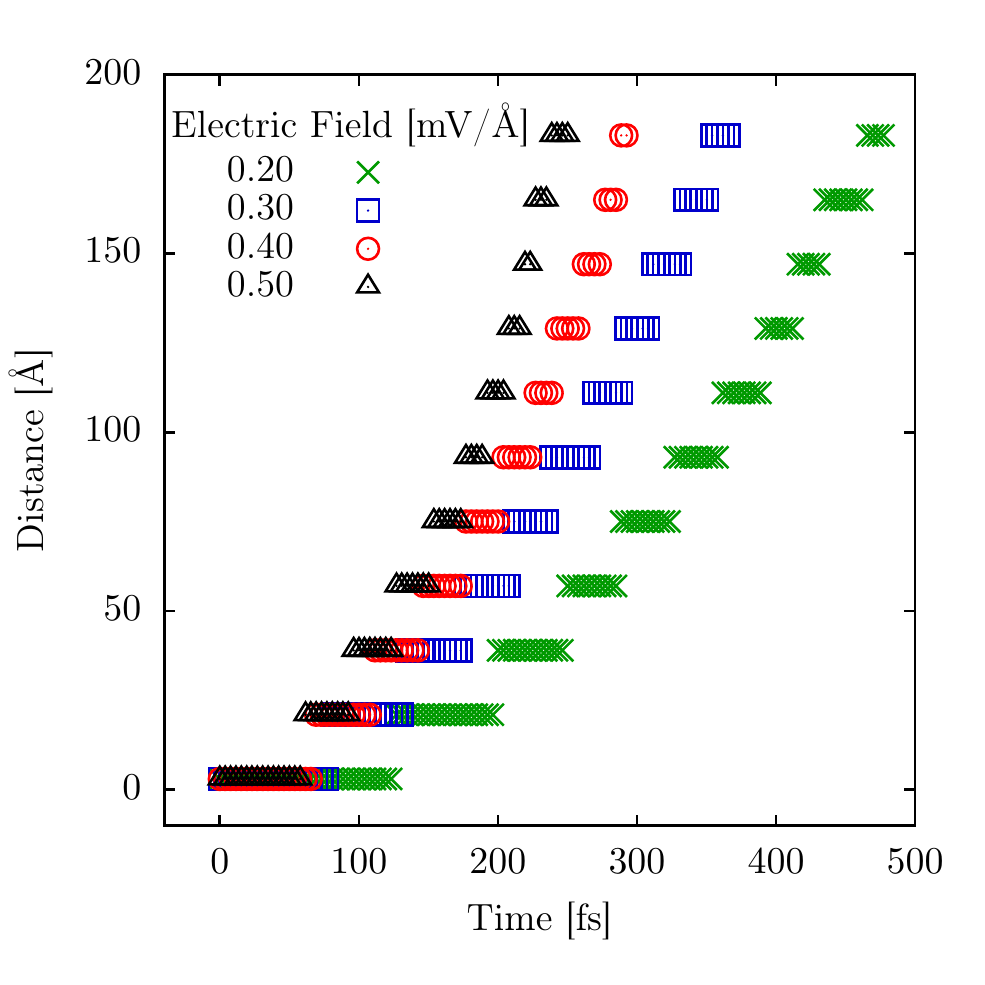}
	\caption{Time-dependent polaron dynamics for 7,9-GNR nanostructure with external electric field presence $ E = 0.3 $ mV/\AA{} (left panel) and time-evolution of the charge center ($x_p$) for different electric field regimes (right panel).}
	\label{fig5}
\end{figure}

This curious preference for polaron formation in the 9-AGNR segments of the system raises the question of whether polarons are able to move in the 7-AGNR regions. In general, for systems in which charge transport is mediated by polarons, the application of an electric field results in charge drift with the center of the quasi-particle moving continuously through the system. In the simulations, the position of the polaron center can be calculated as a function of time by considering the center of the charge distribution as the polaron position ~\cite{junior2019polaron, pereira2019modeling}.

With this assumption at hand, we investigate the motion of a polaron in the 7,9-AGNR under the influence of an electric field. Figure \ref{fig5}(a) shows the behavior of such motion in the case of a 0.3 mV/\AA~ electric field for $\alpha=5.0$ eV/\AA. The same qualitative behavior is observed for polarons as long as $\alpha \ge $ 4.75 eV/\AA. However, as a larger electron-phonon coupling increases the polaron inertia, longer simulations are necessary to observe polaron movement.  It can be seen in Figure \ref{fig5}(a)  that in the first 80 fs of simulation the polaron is mainly localized in  a 9-AGNR segment, with an increase in charge density being observed in the adjacent 9-AGNR portion. This gradual charge transfer is mostly concluded within 100 fs, when the polaron becomes localized in the following 9-AGNR segment. This process is repeated as time increases, but the residence time of the quasi-particle within each 9-AGNR portion is reduced as the polaron gains more momentum. The main aspect of this charge transport process is the discrete nature of the polaron movement between 9-AGNRs as opposed to the expected continuous motion through each alternating AGNR type in the system. This is further corroborated by the behavior of the polaron center, which is shown in Figure \ref{fig5}(b) for electric fields ranging from 0.2 mV/\AA~ to 0.5 mV/\AA. In all cases, the polaron center is seen to remain for a given time in the same position inside a 9-AGNR segment before hopping to the next one. Polaron residence times at each site decrease with time, indicating that the corresponding hopping rate increases. Hopping distances, on the other hand, are kept constant around 15 \AA, which is the distance between neighboring 9-AGNR portions. Finally, compiling results for different intensities of the electric field allows us to estimate charge mobility in the 7,9-AGNR to be $0.144$ cm$^2$/(Vs), which is three orders of magnitude lower than the 350 cm$^2$/(Vs) charge carrier mobility measured for 9-AGNRs~\cite{chen2017chemical}. On the other hand, this constitutes a typical mobility value for organic semiconductors. As such, it is clear that the even though the 7,9-AGNR heterojunction allows for the engineering of bandgpaps in GNRs, it also severely hinders charge mobility in comparison to regular GRNs, restricting polaron motion to a hopping process.

\section{Conclusions}

A semiclassical model with tight-binding approximation was used to describe the structural and electronic properties as well as the charge transport mechanism of an AGNR heterojunction composed of alternating segments of 7-AGNR and 9-AGNR. A sweep of electron-phonon coupling values was conducted, from which its relationship to the energy bandgaps was established. Bond length distortions were presented, indicating that an important structural difference between 7 and 9-AGNR segments was the presence in the latter of aromatic rings. These rings were then shown to suffer strong distortion when a charge was injected to the system, which could be the reason why polarons become localized in the 9-AGNR portions of the system. Furthermore, it was demonstrated that even under the influence of an electric field, the charge carries never localize in the 7-AGNR regions, rather moving by means of a hopping process between 9-AGNRs segments. Hopping rates are seen to increase with time in the initial moments of the simulation but charge mobility reaches only 0.1 cm$^2$/(Vs), which is significantly lower than the experimentally obtained charge mobilities in pure 9-AGNRs but similar to what is found in typical organic semiconductors. We conclude that the engineering of such sequence of heterojunctions in GNRs may allow for gap tuning but simultaneously hinder charge transport in this class of material.

\section*{Acknowledgements}
\noindent The authors gratefully acknowledge the financial support from Brazilian Research Councils CNPq, CAPES, and FAP-DF. L.A.R.J acknowledges the financial support from a Brazilian Research Council FAP-DF and CNPq grants $00193.0000248/2019-32$ and $302236/2018-0$, respectively. 

\bibliographystyle{unsrt}

\bibliography{cas-refs}

\end{document}